\documentclass[twocolumn]{revtex4}
\makeatletter
\def\@hangfrom@section#1#2#3{\@hangfrom{#1#2}{#3}}%
\makeatother
\def\address{\affiliation}
\usepackage[dvips]{graphicx}
\begin{document}

\title{
Transport properties and electronic states of 
the thermoelectric oxide NaCo$_2$O$_4$
}

\author{
I. Terasaki*
}

\address{
Department of Applied Physics, Waseda University, Tokyo 169-8555, Japan\\
Precursory Research for Embryonic Science and Technology,
Japan Science Technology, Tokyo 108-0075, Japan
}

\begin{abstract}
Physical properties of the thermoelectric oxide 
NaCo$_2$O$_4$ are briefly reviewed.
The high thermoelectric properties of this material are 
attributed to the substantially enhanced effective mass,
which comes from the large entropy of Co$^{4+}$ in the low spin state.
The large entropy confined in the CoO$_2$ block
causes a spin-density-wave transition at 22 K
upon Cu substitution for Co, which can be 
regarded as ``order from disorder''.

\end{abstract}

\maketitle
\newpage
\section{Introduction}
Recently layered cobalt oxides have been extensively 
investigated as a promising candidate for a thermoelectric material. 
The thermoelectric material is a material 
that shows large thermopower ($S$),
low resistivity ($\rho$) and low thermal conductivity ($\kappa$)
\cite{mahan}.
Thus far oxides have been regarded as unsuitable for thermoelectric
application because of their poor mobility, but some years ago
Terasaki {\it et al.} found that a single crystal of the layered cobalt oxide
NaCo$_2$O$_4$ exhibits high thermoelectric performance \cite{terra}.
Fujita {\it et al.} showed that 
the dimensionless figure of merit $S^2T/\rho\kappa$ 
of a NaCo$_2$O$_4$ single crystal
exceeds unity at $T=$1000 K \cite{fujita}.
Ohtaki {\it et al.} \cite{ohtaki} measured $S^2T/\rho\kappa \sim 0.8$ at 1000 K
even in the polycrystalline samples of NaCo$_2$O$_4$.
Thus this compound is quite promising for
thermoelectric power generation at high temperature.

We have proposed that the high thermoelectric performance 
of the layered cobalt oxides cannot be explained 
by a conventional band picture based on the one-electron approximation,
but is understood in terms of the strong electron-electron 
correlation effects,
similarly to the case of heavy-fermion compounds.
In fact the material dependence of the thermopower
quite resembles that of Ce$M_2X_2$  ($M$=Cu, Ni, Pd; $X$=Si, Ge)
\cite{terra2}.
Here we briefly review the physical properties of NaCo$_2$O$_4$,
and compare them with those of the heavy-fermion compounds.
We further show anomalous impurity effects \cite{terra3,terra4}, 
which are a prime example of ``order from disorder'' \cite{tsvelik}.

\begin{figure}[b]
 \begin{center}
  \includegraphics[width=7cm,clip]{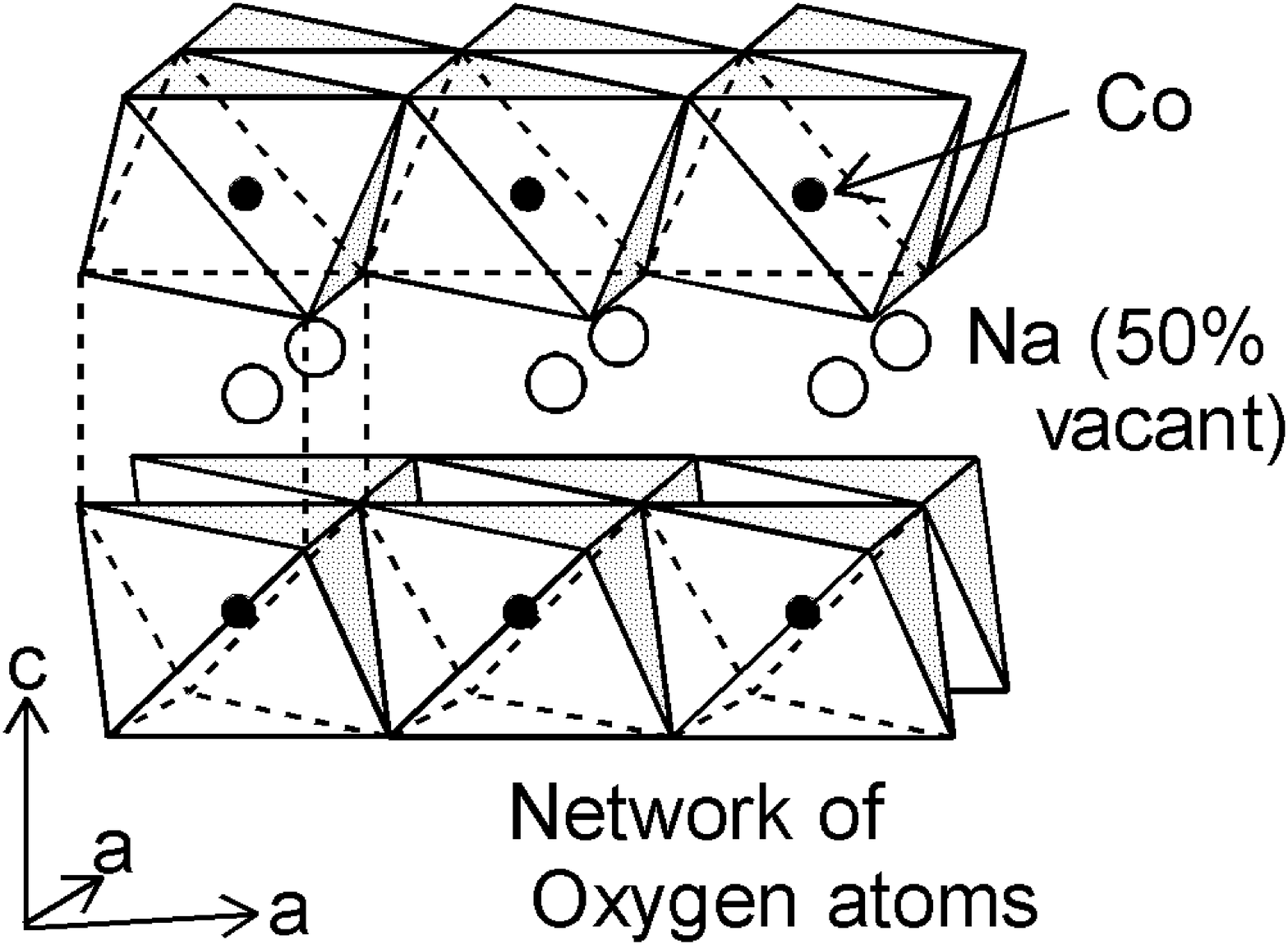}
 \end{center}
\caption{ 
 Crystal structure of NaCo$_2$O$_4$.
 }
\end{figure}

\section{Physical Properties of NaCo$_2$O$_4$}
In Fig.1 is schematically shown the crystal structure of NaCo$_2$O$_4$, 
where edge-shared distorted octahedra of oxygen ions 
form a two-dimensional triangular lattice. 
A cobalt ion is in the center of the distorted octahedra 
to form a two-dimensional triangular lattice, 
while a sodium ion is in a prism site between CoO$_2$ blocks. 
An important feature in this structure is that 
sodium ions randomly occupy the regular site by 50\%, 
and the sodium content changes from 50 to 75\%. 
In this sense NaCo$_2$O$_4$ should be written as Na$_x$CoO$_2$
($x$=0.5). 
Nevertheless we will call it NaCo$_2$O$_4$, because the best thermoelectric 
properties are realized near the 50\% Na occupancy \cite{ohtaki}.

\begin{figure}
 \begin{center}
  \includegraphics[width=8cm,clip]{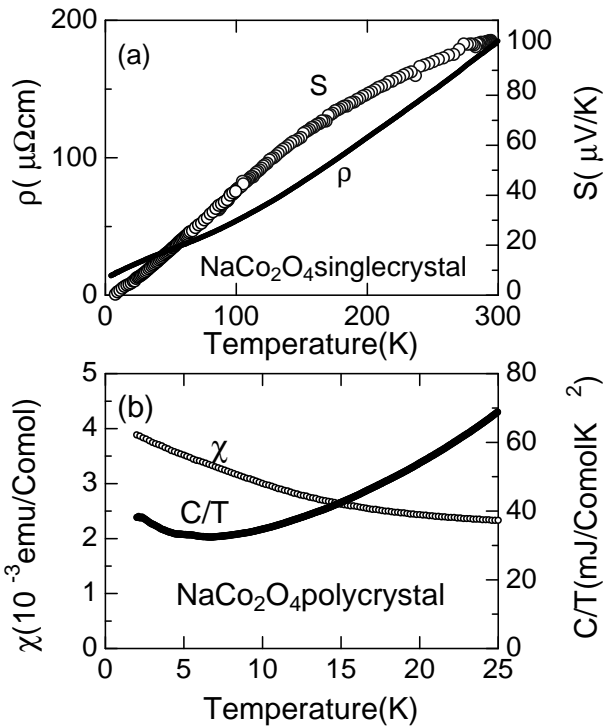}
 \end{center}
 \caption{ 
 (a) Resistivity ($\rho$) and thermopower ($S$) of 
 a NaCo$_2$O$_4$ single crystal along the in-plane direction;
 (b) Susceptibility ($\chi$) and specific heat ($C$)
 of NaCo$_2$O$_4$ polycrystals.
 }
\end{figure}

Figure 2(a) shows the resistivity and the thermopower of a NaCo$_2$O$_4$ 
single crystal along the a-axis (parallel to the CoO$_2$ block) direction. 
The magnitude of the resistivity is as low as 200 $\mu\Omega$cm at 300 K, 
which corresponds to nearly the best conductivity 
among transition-metal oxides. 
The thermopower is 100 $\mu$V/K at 300 K, 
which is as large as that of degenerate semiconductors. 
As a result, the power factor $S^2/\rho$ is as large as 
(or even larger than) $S^2/\rho$ of the state-of-the-art 
thermoelectric material Bi$_2$Te$_3$ \cite{terra}.

Figure 2(b) shows the magnetic susceptibility $\chi$
and the specific heat $C$ of a polycrystalline 
sample of NaCo$_2$O$_4$ \cite{terra4,ando}. 
The susceptibility is relatively large 
(4$\times$10$^{-3}$ emu/Co mol at 4 K), 
and weakly temperature dependent. 
We attribute the weakly temperature dependent susceptibility 
to a kind of spin fluctuation, 
as is seen in the valence fluctuation systems. 
In fact, Co-site substitution causes non-trivial change 
in the susceptibility. 
Nonmagnetic Zn substitution induces a Curie term, 
whereas magnetic Cu substitution decreases 
the magnitude of the susceptibility, 
which cannot be interpreted by the local 
spin picture \cite{cimtec}.

The electron specific heat coefficient $\gamma$ is 
evaluated by the value of $C/T$ as $T\to$0. 
$C/T$ approaches 35-50 mJ/Co molK$^2$, 
which is two orders of magnitude larger than the value of conventional metals. 
The large values of $\gamma$ and $\chi$ clearly indicate 
that the density of states (and also the effective mass) 
is substantially enhanced in NaCo$_2$O$_4$.
Note that the band calculation by Singh \cite{singh}
also predicts the large density of states,
though it is difficult to explain the anomalous 
impurity effects discussed later.

\section{Comparison with Ce-based intermetallics}
The microscopic theory for the high thermoelectric performance 
of NaCo$_2$O$_4$ is still lacking, but we think that the following features 
are now established.
(1) The mixture of Co$^{3+}$ and Co$^{4+}$ in the low spin state
can carry a large entropy of $k_B$log6, which gives a thermopower
of 150 $\mu$V/K in the high-temperature limit \cite{koshibae}.
(2) NaCo$_2$O$_4$ shows no structural, electric, and magnetic
transitions from 2 to 1000 K.
(3) From (1)(2), the large entropy cannot be released
through phase transitions, and inevitably point to
the conducting carriers to form a ``heavy-fermion''-like 
electron \cite{terra4}.
(4) Although an exact one-to-one correspondence is not realized,
the calculated Fermi surface of NaCo$_2$O$_4$ is made from
two orbitals of different nature \cite{singh},
just like Ce $4f$ and $sp$ bands in the Ce-based intermetallics 
\cite{terra2,terra4}.

In the heavy-fermion compounds, $\gamma$ is proportional 
to the susceptibility $\chi$, and $\chi/\gamma$  
(the Wilson ratio) is independent of materials. 
In Fig.3, various $\gamma$'s are plotted as a function of $\chi$, 
where the data for NaCo$_2$O$_4$ is just close to the data 
for the valence-fluctuation compound CePd$_3$ \cite{ict1999}. 
We should note that CePd$_3$ has nearly the same values of $S$ 
(80 $\mu$V/K at 300 K) and $\rho$ (150 m$\Omega$cm at 300 K). 
Unfortunately it has much higher thermal conductivity 
than NaCo$_2$O$_4$, 
which greatly reduces the thermoelectric performance.

\begin{figure}
 \begin{center}
  \includegraphics[width=8cm,clip]{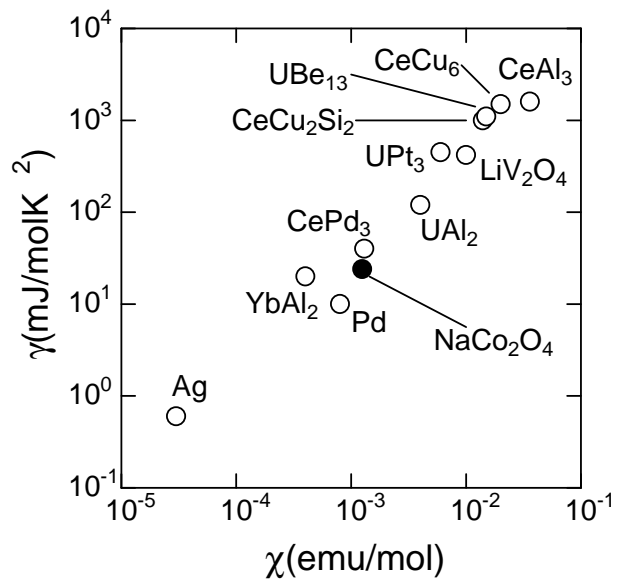}
 \end{center}
 \caption{ 
 Electron specific heat coefficient $\gamma$
 as a function of magnetic susceptibility ($\chi$)
 for various correlated metals.
 }
\end{figure}

According to the classical Boltzmann theory, 
the diffusive term of the thermopower of a metal is understood 
as transport entropy per carrier,  implying that 
a large effective mass can lead to a large thermopower. 
Figure 3 indicates that the effective mass of NaCo$_2$O$_4$
is more than 10 times larger than the bare electron mass, 
which is in a quantitative agreement with the reflectivity edge 
and the magnitude of the Hall effect \cite{ict1999}. 
Considering a striking similarity to the heavy fermion compounds, 
we attribute the large mass of NaCo$_2$O$_4$ to the coherent state 
between the conduction carrier and the spin fluctuation.
Very recently Valla {\it et al.} have performed a photoemission 
experiment for the layered cobalt oxides,
and have successfully observed the coherent state 
at low temperatures \cite{valla}.

\begin{figure}
 \begin{center}
  \includegraphics[width=8cm,clip]{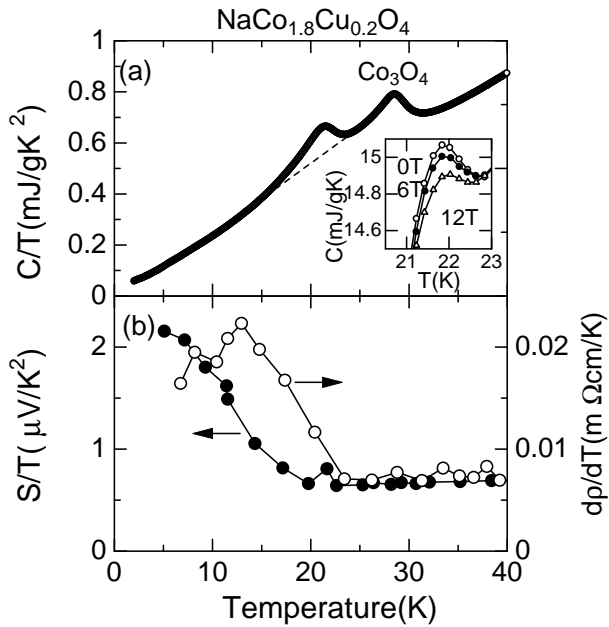}
 \end{center}
 \caption{ 
 (a) Specific heat coefficient $C$
 (b) Temperature derivative of the resistivity
 ($d\rho/dT$) and temperature coefficient of the thermopower $S/T$
  for the Cu-substituted NaCo$_{1.8}$Cu$_{0.2}$O$_4$.
 Inset: The specific heat in magnetic fields.}
\end{figure}

\section{Cu substitution effects: Order from disorder}
Figure 4(a) shows the specific heat for 
the Cu-substituted sample NaCo$_{1.8}$Cu$_{0.2}$O$_4$
as a function of temperature,
where the specific heat jump at 22 K is clearly seen \cite{terra4}.
The entropy change of the 22-K transition is as small as 
77 mJ/Kmol, corresponding to 0.01$k_B$ per Co.
Actually only 5\% of Co$_3$O$_4$ impurity 
exhibits a specific heat jump of the same order at 30 K.
A phase transition accompanied by a small entropy change
is an off-diagonal long range order,
which induces a small entropy change of the order of $Nk_BT/E_F$.
This order is suppressed by external fields, as shown in the 
inset of Fig. 4(a).

Figure 4(b) shows the $T$-linear term of the thermopower ($S/T$)
and the temperature derivative of the resistivity $d\rho/d T$.
Their temperature dependences are quite similar to each other,
where the magnitude increases up to almost twice below 22 K.
This indicates that the Drude weight decreases by 50\% at low
temperatures, implying the existence of a (pseudo)gap.
As an off-diagonal long-range order with a gapped state,
one would think of charge density wave (CDW) or spin density wave (SDW).
The calculated Fermi surface 
of the $a_{1g}$ band is hexagon-like \cite{singh}, 
which is unstable against CDW or SDW formation with the nesting vector
along the $\Gamma$-K direction.
We think that the 22-K transition is SDW-like,
because CDW is insensitive to magnetic field.
Actually we can find many similarities between the 22-K transition and 
the SDW transition in Cr \cite{CrReview}. 

NaCo$_2$O$_4$ is very close
to the instability for various phase transitions arising 
from the large entropy per site.
The Cu substitution enhances the instability,
and eventually causes the SDW-like transition at 22 K.
This type of transition is called ``order from disorder''
\cite{tsvelik}.
In other words, instabilities against various phases are competing 
or disordering in NaCo$_2$O$_4$,
and any phase transitions are prohibited down to low temperatures.
This does not mean that NaCo$_2$O$_4$ is far from 
the instability of phase transitions, but rather, 
is very susceptible to various transitions against various perturbations.
In fact, Na$_{1.5}$Co$_2$O$_4$ exhibits 
a spin-glass behavior at 3 K \cite{takeuchi}
and a ferromagnetic transition at 22 K \cite{motohashi},
possibly owing to the structure instability of the $\gamma$ phase,
and (Bi,Pb)-Sr-Co-O shows a ferromagnetic transition at 4 K 
due to the lattice misfit \cite{tsukada}.

Let us qualitatively discuss 
why the SDW-like state is favored in NaCo$_{2-x}$Cu$_x$O$_4$.
SDW and CDW are closely related to the nesting of the 
and the topology of the Fermi surface, and 
occur when the correlation effect is weak enough
to hold one-electron picture based on the band calculation.
We think that Cu suppresses the mass enhancement 
without significant change in the carrier concentration \cite{terra4}.
If so, the substituted Cu enhances the screening of 
the magnetic fluctuation, 
which might recover the band picture to cause
the CDW/SDW instability of the $a_{1g}$ Fermi surface.

\section{Summary}
We have reviewed the anomalous properties of NaCo$_2$O$_4$, 
which are promising for thermoelectric applications. 
We have explained the thermoelectric properties of NaCo$_2$O$_4$ 
in terms of the layered structure consisting of 
the strongly correlated CoO$_2$ layer.
Carries in oxides are often coupled with the optical phonons, 
the spin degrees of freedom and the orbital degrees of freedom. 
These couplings induce exotic electronic states such as polarons, 
heavy fermions, spin liquids and orbital liquids. 
Furthermore, the geometry and the dimensionality are easy 
to change in certain classes of oxides, 
where layered, ladder and chain structures 
can be controlled. 
Therefore I believe that many functional materials 
including thermoelectric materials will still sleep unknown. 
I hope that NaCo$_2$O$_4$ is just the beginning, 
and that a thermoelectric oxide of higher performance 
will appear in near future.

I would like to thank I. Tsukada, T. Motohashi and H. Yamauchi 
for collaboration, and also thank S. Kurihara and Y. Tsunoda 
for fruitful discussion of spin density wave.


\begin{thebibliography}{99}
 \bibitem{mahan}
	 G. D. Mahan:
	 Solid State Physics 51 (1998) 81.
 \bibitem{terra}
	 I. Terasaki, Y. Sasago, and K. Uchinokura:
	 Phys. Rev. B56 (1997) R12685.
 \bibitem{fujita}
	 K. Fujita, T. Mochida, and K. Nakamura:
	 Jpn. J. Appl. Phys. 40 (2001) 4644.
 \bibitem{ohtaki}
	 M. Ohtaki, Y. Nojiri and E. Maeda:
	 {\it Proc. of the 19th International 
	 Conference on Thermoelectrics (ICT2000)},
	 (Babrow, Wales, 2000) p.190. 
 \bibitem{terra2}
	 I. Terasaki:
	 Mater. Trans. 42 (2001) 951.
 \bibitem{terra3}
	 I. Terasaki, Y. Ishii, D. Tanaka, K. Takahata, 
	 and Y. Iguchi:
	 Jpn. J. Appl. Phys. 40 (2001) L65.
 \bibitem{terra4}
	 I. Terasaki, I. Tsuakda and Y. Iguchi:
	 Phys. Rev. B, 65 (2002) 195106.
 \bibitem{tsvelik}
	 A. M. Tsuvelik:
	 ``Quantum field theory in condensed matter physics'',
	 (Cambridge University Press, 1995, Cambridge) p.174.
 \bibitem{ando}
	 Y. Ando,  Y. Hanaki, S. Ono, T. Murayama, 
	 K. Segawa, N. Miyamoto and S. Komiya:	
	 Phys. Rev. B61 (2000) R14956. 
 \bibitem{cimtec}
	 I. Terasaki:
	 {\it Proc. of International Conference on Mass
          and Charge Transport in Inorganic Materials} 
          (Techna Srl, 2000) p.1333
 \bibitem{singh}
	 D. J. Singh:
	 Phys. Rev. B61 (2000) 13397. 
 \bibitem{koshibae}
	 W. Koshibae, K. Tsutsui and S. Maekawa:
	 Phys. Rev. B 62 (2000) 6869.
 \bibitem{ict1999}
	 I. Terasaki: {\it Proc. of the 18th International 
	 Conference on Thermoelectrics (ICT 1999)} 
	 (IEEE, Piscataway, 2000), p.569
 \bibitem{valla}
	 T. Valla, P. D. Johnson, Z. Yusof, B. Wells, Q. Li,
	 S. M. Loureiro, R. J. Cava, M. Mikami, Y. Mori, M. Yoshimura
	 and T. Sasaki:
	 Nature 417 (2002) 627.
 \bibitem{CrReview}
	 E. Fawcett, H. L. Alberts, V. Yu. Galkin, D. R. Noakes, and
	 J. V. Yakhmi:
	 Rev. Mod. Phys. 66 (1996) 25.
 \bibitem{takeuchi}
	 T. Takeuchi, M. Matoba, T. Aharen, and M. Itoh:
	 Physica B 312-313 (2002) 719.
 \bibitem{motohashi}
	 T. Motohashi{\it et al.}:
	 submitted to Phys. Rev. Lett.
 \bibitem{tsukada}
	 I. Tsukada, T. Yamamoto, M. Takagi, T. Tsubone, S. Konno, 
	 and K. Uchinokura:
	 J. Phys. Soc. Jpn. 70 (2001) 834.
\end{thebibliography}
\end{document}